\documentclass[11pt,titlepage]{article}
\pdfoutput=1
\usepackage{amsmath}
\usepackage{amssymb}

\usepackage[pdftex]{graphicx}
\usepackage[labelfont=bf,labelsep=period,justification=raggedright]{caption}

\usepackage{cite}
\DeclareGraphicsExtensions{.pdf,.png,.jpg,.eps}
\usepackage[usenames,dvipsnames]{xcolor} 
\usepackage{epstopdf}

\makeatletter
\renewcommand{\@biblabel}[1]{\quad#1.}
\makeatother

\date{}

\newcommand{\K}{{\textrm{K}^+}}
\newcommand{\Na}{{\textrm{Na}^+}}
\newcommand{\OO}{{\textrm{O}_2}}
\newcommand{\Cl}{{\textrm{Cl}^-}}
\newcommand{\mM}{{\textrm{mM}}}
\newcommand{\mV}{{\textrm{mV}}}

\newcommand{\ATP}{{\textrm{ATP}}}
\newcommand{\AMP}{{\textrm{AMP}}}
\newcommand{\ADP}{{\textrm{ADP}}}

\begin{document}
\begin{flushleft}
{\Large
\textbf{A mathematical model of the metabolic and perfusion effects on cortical spreading depression}
}
\\
Joshua C. Chang$^{1\ast}$, 
K.C. Brennan$^{2}$, 
Dongdong He$^{3}$,
Huaxiong Huang$^{4}$,
Robert M. Miura$^{5}$,
Phillip L. Wilson$^{6}$,
Jonathan J. Wylie$^{7}$
\\
\textbf{1} Mathematical Biosciences Institute, The Ohio State University, Columbus, Ohio, USA. {\tt chang.1166@mbi.osu.edu}
\\
\textbf{2} Department of Neurology, University of Utah, Salt Lake City, Utah, USA. {\tt k.c.brennan@hsc.utah.edu}
\\
\textbf{3} Department of Mathematics, City University of Hong Kong, Kowloon Tong, Hong Kong. {\tt dongdohe@cityu.edu.hk}
\\
\textbf{4} Department of Mathematics and Statistics, York University, Toronto, Ontario, Canada M3J 1P3. {\tt hhuang@yorku.ca}
\\
\textbf{5} Department of Mathematical Sciences and Center for Applied Mathematics and Statistics, New Jersey Institute of Technology, Newark, NJ 07102 USA. {\tt miura@njit.edu} 
\\
\textbf{6} Department of Mathematics and Statistics, University of Canterbury, Christchurch, New Zealand. {\tt  p.wilson@math.canterbury.ac.nz}
\\
\textbf{7} Department of Mathematics, City University of Hong Kong, Kowloon Tong, Hong Kong. {\tt mawylie@cityu.edu.hk}\\
$\ast$ E-mail: {\tt chang.1166@mbi.osu.edu}
\end{flushleft}

\abstract{
Cortical spreading depression (CSD) is a slow-moving ionic and metabolic disturbance that propagates in cortical brain tissue. In addition to massive cellular depolarization, CSD also involves significant changes in perfusion and metabolism -- aspects of CSD that had not been modeled and are important to traumatic brain injury, subarachnoid hemorrhage, stroke, and migraine.

 In this study, we develop a mathematical model for CSD where we focus on modeling the features essential to understanding the implications of neurovascular coupling during CSD. In our model, the sodium-potassium--ATPase, mainly responsible for ionic homeostasis and active during CSD, operates at a rate that is dependent on the supply of oxygen. The supply of oxygen is determined by modeling blood flow through a lumped vascular tree with an effective local vessel radius that is controlled by the extracellular potassium concentration. We show that during CSD, the metabolic demands of the cortex exceed the physiological
limits placed on oxygen delivery, regardless of vascular constriction or dilation. However,
vasoconstriction and vasodilation play important roles in the propagation of CSD and its recovery.  Our model replicates the qualitative and quantitative behavior of CSD -- vasoconstriction, oxygen depletion, extracellular potassium elevation, prolonged depolarization -- found in experimental studies.
  
  We predict faster, longer duration CSD in vivo than in vitro due to the contribution of the vasculature. Our results also 
help explain some of the variability of CSD between species and even within the same animal.
 These results have clinical and translational implications, as they allow for more precise in vitro, in vivo, and in silico exploration of a phenomenon broadly relevant to neurological disease.

}

\section*{Author Summary}

Cortical spreading depression (CSD) is a traveling wave of depolarization that is associated with migraine and brain injury. Massive changes in cerebral blood flow have been associated with CSD, but thus far no mathematical modeling has incorporated these physiologically important changes. In this manuscript, we present a mathematical model for CSD that incorporates blood flow and oxygen consumption in order to examine the effects of the brain vasculature on CSD and vice versa.


\section{Introduction}\label{sec:intro}

Cortical spreading depression (CSD) is a self-propagated depolarization that occurs in the gray matter of many species~\cite{bures1974mechanism}. In humans, it is known to occur during brain injury, stroke, and subarachnoid hemorrhage \cite{kraig2002spreading}. There is also strong evidence that CSD is responsible for the migraine aura \cite{tfelt-hansen2010migraine-csd,hadjikhani2001mechanisms,lashley1941patterns}, a sensory hallucination associated with the migraine attack.

Although CSD was discovered in 1944 by Le\~ao \cite{le1944spreading}, we still do not have a detailed understanding of how CSD is manifest.  In particular, CSD has been associated with massive changes in cortical perfusion. The magnitudes of these changes vary by animal species, but significant decreases and increases in blood flow volume occur in all species tested~\cite{brennan2010update,charles2009cortical,busija2008mechanisms}. Also common to all species tested is a mismatch in the delivery of substrates to meet metabolic demands, resulting in a derangement of neurovascular coupling~\cite{chang2010biphasic,piilgaard2009persistent}. 

Blood delivery is known to play a significant role in CSD in several ways.  Changes in perfusion can induce peri-infarct depolarizations (PID), which are electrophysiologically identical to CSD. Conditions that mimic the effects of hypoperfusion, such as oxygen glucose deprivation (OGD) and exposure to ouabain (an inhibitor of the $\Na/\K$--ATPase), also generate spreading depolarizations~\cite{somjen2001mechanisms,lauritzen2010clinical,dreier2011role}.


Perfusion changes can also modulate the signature characteristics of CSD. Depending
on the levels of the underlying oxygenation or blood pressure, the amplitude and duration of
depolarization and the velocity of propagation of CSD can be altered~\cite{dreier2011role,sukhotinsky2008hypoxia}. 
Clearly, CSD in vivo cannot be understood without reference to the vascular
changes that condition -- and are conditioned by -- the phenomenon.  There is a need to further explore the implications of the effects of perfusion and metabolism on various aspects of CSD.  

Such explorations naturally lead to the development of mathematical models in which many mechanisms can be studied independently and/or simultaneously.  Previous mathematical models of CSD have accounted for ionic diffusion~\cite{tuckwell1978mathematical}, cellular membrane ionic currents~\cite{tuckwell1978mathematical,kager2000simulated}, the $\Na/\K$--ATPase and other membrane pumps~\cite{tuckwell1978mathematical,kager2000simulated},  and extra- and intracellular volume changes~\cite{shapiro2001osmotic,yao2010continuum}. 

However, mathematical models of CSD have not looked at the dynamical implications of neurovascular coupling and metabolism.  Here we formulate a five-compartment continuum model for CSD  that uses known physiological data relating effective blood vessel diameter and extracellular potassium concentration to model oxygen delivery in the brain.  The compartments in our model are:  a somatic  neuronal compartment, a dendritic neuronal compartment, an extracellular space compartment, a vascular tree compartment, and a glial compartment.

We show that the oxygen deprivation that results from both metabolic demand and vasoconstriction modifies the characteristics of CSD waves. Our results predict faster, longer duration CSD \emph{in vivo} than \emph{in vitro}, due to the contribution of the vasculature. Our results also help explain some of the variability of CSD between species and even within the same animal.   In addition, the model explains differences between CSD \emph{in vivo} and CSD in brain slices due to variant arterial constriction and dilation during the CSD event.

\section{Methods\label{sec:model}}

\begin{figure}[t]
\includegraphics[width=\textwidth]{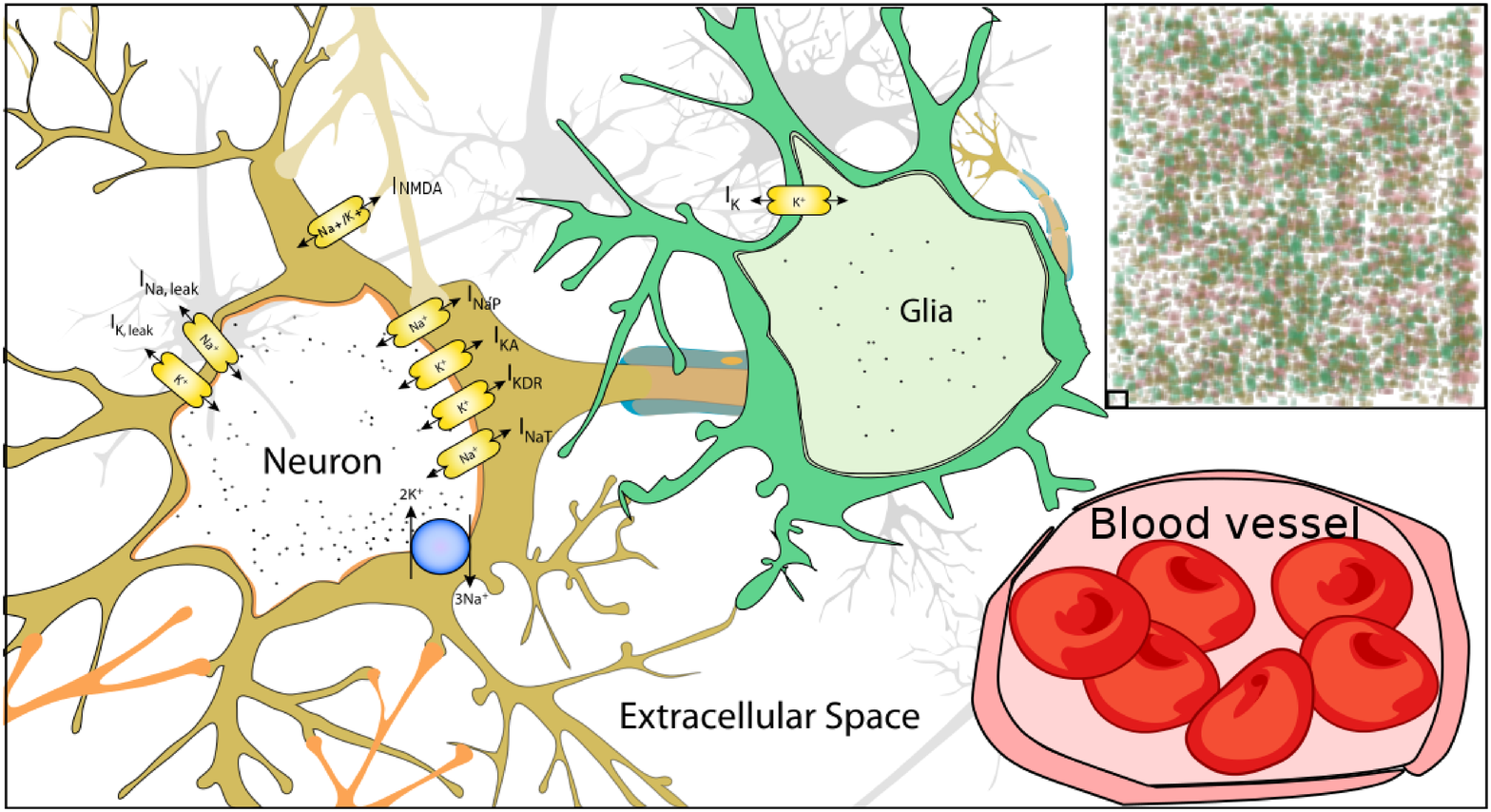}
\caption[{\textbf{Diagram of the model.} Neurons (mustard) consist of various sodium and potassium channels as well as the {$\Na/\K$--}ATPase. Dendrites additionally consist of NMDA channels. Glia (teal) are incorporated as potassium buffers. Blood vessels (pink) bring in oxygen to supply the {$\Na/\K$--}ATPase. The cellular-level model is taken into the continuum limit (upper right) to yield a model with 5 compartments: neural cell bodies, neural dendrites, glial, vascular, and extracellular space. }]{}
\label{fig:diagram}
\end{figure}

To study the important new elements that affect and are affected by CSD, we formulate a five-compartment continuum mathematical model, see Fig~\ref{fig:diagram}. Neurons comprise two of the five compartments: a compartment representing the dendritic processes ($d$), and a compartment representing the cell bodies (somatic compartment $s$). The {ECS} ($e$), the vascular bed ($v$), and glia {(g)} comprise the remaining compartments. Though cell swelling has been shown to occur during CSD~\cite{takano2007cortical,andrew2007physiological}, we make the simplifying assumptions that the {ICS and ECS} volume fractions remain fixed, based on the findings of Yao et al.~\cite{yao2010continuum} who found that osmotic effects do not significantly affect the propagation of CSD waves.



The {ICS and ECS} compartments include only the {most relevant} ions (sodium, potassium, chloride) and channels that have been shown to be responsible for the instigation and spread of CSD~\cite{yao2010continuum,kager2000simulated,kager2002conditions}. In the somatic membranes, we include T-type sodium channels, P-type sodium channels, delayed-rectifier potassium channels,  A-type potassium channels, and the $\Na/\K$--ATPase. In the dendritic membranes, we additionally include NMDA channels.

Our model assumes that the vascular compartment does not exchange fluid with the extracellular space. The effective diameters of proximal arterioles {control} the blood flow rate.  In turn, the vascular diameters are coupled to neuronal activity through {ECS} potassium concentrations proximal to dendritic processes, which is also buffered by astrocytes.   

\subsection{Membrane potential and ion transport using a neuronal model}\label{sec:membraneion}
The membrane potentials  of the neuronal compartments, $E_{m,\ast}$ ($\ast$ is either $s$ for somatic or $d$ for dendritic), are governed by the coupled partial differential equations
\begin{align}
C_m\frac{\partial E_{m,s}}{\partial t}&=-I_{s,\textrm{tot}} +\overbrace{\frac{1}{2R_a\delta_d^2}\left(E_{m,d}-E_{m,s}\right)}^{\textrm{dendrite coupling}}{,} \\
C_m\frac{\partial E_{m,d}}{\partial t}&=-I_{d,\textrm{tot}} +\overbrace{\frac{1}{2R_a\delta_d^2}\left(E_{m,s}-E_{m,d}\right) }^{\textrm{soma coupling}} 
\end{align}
where $C_m$ is the membrane capacitance per unit surface area ($\mu$farad/cm$^2$) for both the somatic and dendritic membranes, $R_a$ is the input resistance of the effective dendritic tree (ohms), $\delta_d$ is the half length of the effective dendritic tree (cm), and $\displaystyle I_{\ast,\textrm{tot}}=\sum_{\textrm{ions}} I_{\ast,\textrm{ion,tot}}=\sum_{\textrm{ions}} \sum_{\textrm{channels}}I_{\ast,\textrm{ion,channel}}$ are the spatially-dependent total cross-membrane ionic currents per unit surface area (mA/cm$^2$) for each neuronal compartment. Following Kager et al.~\cite{kager2000simulated,kager2002conditions} and Yao et al.~\cite{yao2010continuum}, the total cross-membrane currents, $I_{\ast,\textrm{tot}}$, are given for the three major ions (sodium, potassium, and chloride) and are the sum of the active and passive (leak) sodium and potassium currents, the chloride (leak) current, and the sodium-potassium exchange pump current (see Appendix \ref{appendix}).  Here we model the ion exchange between the somatic and dendritic portions of the neuron by a flux proportional to the difference between the ion concentrations. The exchange coefficient $D_{ion}/2\delta^2_d$ is estimated using the molecular diffusion coefficient and mean length of the dendrites, adjusted by the volume ratio of soma and dendrites.

The local rates of change of the ECS ions ($\Na, \K$, and $\Cl$) are due to membrane ionic currents, diffusion of extracellular ions, and the buffering of ECS potassium by glial cells. Note that all of the model differential equations have only time derivatives in them with the exception of the ECS diffusion equations for the ions. However, all of these equations depend implicitly on the spatial coordinate as a result of the spatial distribution of the ions.
\begin{align}
\frac{\partial([\textrm{ion}]_{e})}{\partial t}&=\overbrace{\frac{1}{f_eF}\left\{\frac{{A}_sI_{s,\textrm{ion,tot}}}{ V_s}+\frac{{A}_dI_{d,\textrm{ion,tot}}}{ V_d} \right\}}^{\textrm{through channels}}
+\overbrace{\frac{\partial}{\partial x}\left(D_{\textrm{ion}}\frac{\partial [\textrm{ion}]_{e}} {\partial x}\right)}^{\textrm{extracellular diffusion}}.  \label{eq:ecs} 
\end{align}
The notation, $D_\textrm{ion}$, corresponds to the ion diffusion coefficient in aqueous solution taking into account tortuosity and volume fraction, see~\cite{nicholson1981ion,nicholson2001diffusion}, and $F$ is the Faraday constant. The quantities
 ${A}_{(\ast)}$ are the surface areas of the neuronal compartments in the total fixed volume given by the sum of the fixed somatic volume $V_s$, dendritic volume $V_d$, and extracellular volume, $V_e$.   The ECS volume fraction is given by $f_e=V_e/(V_s+V_d)$.  The equation for {ECS} potassium is modified by adding the buffering flux term, $v_\textrm{buffer}$, given in Section~\ref{sec:buffer}.  The equations for the rates of change of {ICS} ions are
\begin{align}
\frac{\partial([\textrm{ion}]_{i,s})}{\partial t}&=\overbrace{-\frac{{A_s}}{FV_s}I_{s,\textrm{ion,tot}}}^{\textrm{through channels}} + \overbrace{\frac{D_{\textrm{ion}}(V_d+V_s)}{2\delta_d^2V_s}\left([\textrm{ion}]_{i,d}-[\textrm{ion}]_{i,s} \right) }^{\textrm{exchange between soma and dendrites}}\label{eq:ics1} \\
\frac{\partial([\textrm{ion}]_{i,d})}{\partial t}&=-\frac{{A_d}}{FV_d}I_{d,\textrm{ion,tot}}+\frac{D_{\textrm{ion}}(V_s+V_d)}{2\delta_d^2V_d}\left([\textrm{ion}]_{i,s}-[\textrm{ion}]_{i,d} \right) . \label{eq:ics2}
\end{align}
Note that we have not included the diffusion of {ICS} ions because any appreciable diffusion (relative to cell size) would require {an ion} to first become an ECS ion, diffuse extracellularly, and then becomes an ICS ion. 

As Cl$^-$ is the only ECS anion in our simulations, to ensure electroneutrality, its initial extracellular concentration ($[\Cl]_{e}$) was determined by the sum of the concentrations of the major cations: $[\Cl]_{e}=[\Na]_{e}+[\K]_{e}$.  We chose the initial intracellular concentrations of chloride ($[\Cl]_{i,\ast}$) so that its Nernst potentials matched the membrane potentials. To achieve intracellular electroneutrality  in the soma and dendrites, we assume the existence of immobile anions.  Although calcium can have major effects on neurotransmitter dynamics and other secondary messenger effects, we ignored the calcium concentration because of its relatively small values.

\subsection{Exchange pumps}

During CSD, the ionic concentrations in the {ECS} and {ICS} are considerably displaced from steady-state. This displacement occurs primarily because of fluxes through voltage-gated sodium and potassium channels.  Here, we include the sodium-potassium exchange pumps ({$\Na/\K$}--ATPase) in the neuronal membranes, whose primary role is to restore the ionic concentrations back to their homeostatic state.  {The ionic pumps are active and consume energy.}  When local oxygen levels are depleted, ATP is in short supply.  Therefore, the function of the ionic pumps {in our model} is related to oxygen consumption and vascular flows.  

These pumps are involved in the movements of  {ICS} sodium and {ECS} potassium against their electrochemical gradients {and} require active ionic pumps that consume {energy.  The} pumps are fueled by the dephosphorylation of ATP in the cell~\cite{keener2009mathematical} given by 
$$\ATP + 3{\Na}_i+2{\K}_e \xrightarrow{\textrm{pump}} \ADP + P_i+3{\Na}_e+2{\K}_i.$$
ATP is replenished by the reattachment of a phosphate ion to ADP {and} is powered by cellular respiration through both aerobic and anaerobic processes. When local oxygen in the tissue is very low, the normal ATP dynamics are perturbed.

The $\Na/\K$--ATPase is a trans-membrane protein with two extracellular binding sites for potassium, three intracellular binding sites for sodium, and a single intracellular binding site for ATP.  In each neuronal  compartment, its potassium and sodium currents are given by $I_{\ast,\textrm{K,pump}}=-2I_{\ast,\textrm{pump}}$ and $I_{\ast,\textrm{Na,pump}}=3I_{\ast,\textrm{pump}}$  (as noted in the beginning of Section~\ref{sec:membraneion}, $\ast$ is either $s$ for somatic or $d$ for dendritic), respectively, with \begin{equation}
\displaystyle I_{\ast,\textrm{pump}} = I_\textrm{max}\gamma_{\ast,\textrm{pump},1}{\gamma_{\textrm{pump},2}}\label{eq:pump}\nonumber
\end{equation} 
where
\begin{align}
\gamma_{\ast,\textrm{pump},1}([\K]_{e},[\Na]_{i,\ast})
=\left(1+\frac{[\K]_{e,0}}{[\K]_{e}}\right)^{-2}\left(1+\frac{[\Na]_{i,0}}{[\Na]_{i,\ast}}\right)^{-3}.\label{eqn:pumpnak}
\end{align}
Equation~\ref{eqn:pumpnak} is given in~\cite{kager2000simulated} and the expressions on the right are dependent on {ECS} potassium and {ICS} sodium concentrations. These equations implicitly assume that ATP is plentiful, which is not the case when oxygen is limited or when metabolic needs are high. 

\begin{figure}
\begin{center}
\includegraphics[width=19.62pc]{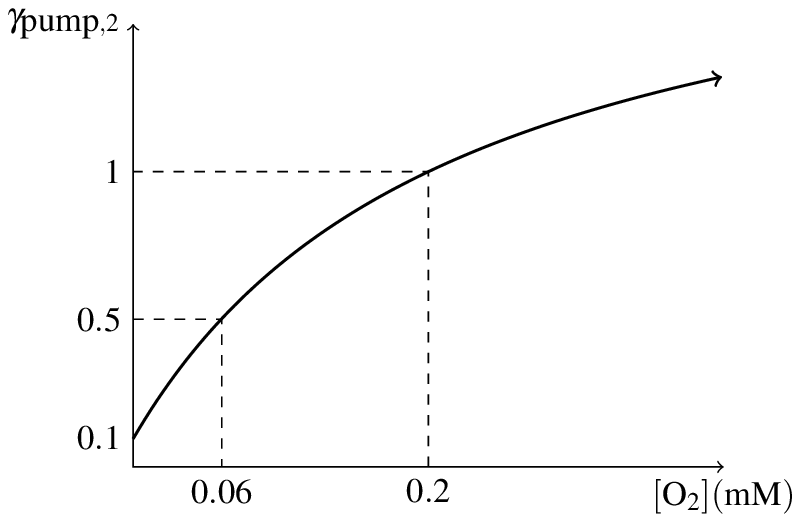}
\caption[{\textbf{Oxygen availability affects the $\Na/\K$--ATPase.} Shown is the relationship between tissue oxygen concentration and the oxygen-dependent portion of the pump rate (Eq~\ref{eqn:gammaO}). Since some ATP is generated even in the absence of oxygen, the pump rate does not completely go to zero as the oxygen concentration approaches zero. The steady state oxygen concentration is $0.2 \mM$.}]{}
\label{fig:gammaO}
\end{center}
\end{figure}

In the oxygen-limited regime, where ATP is limited, we modify this expression with an
additional oxygen {dependent} term
\begin{equation}
\gamma_{\textrm{pump},2}([\OO])=2\left(1+\frac{[\OO]_0}{(1-\alpha)[\OO]+\alpha[\OO]_0}\right)^{-1}
\label{eqn:gammaO}
\end{equation}
where $\alpha$ is the percentage (about 5\%) of ATP production that is independent of oxygen~\cite{klein2003principles,chih2003energy},  the subscript $0$ denotes the equilibrium values, and $[\OO]$ is the tissue oxygen concentration (see Fig~\ref{fig:gammaO}). This expression indicates that the pumping rate will be reduced whenever there is a decrease of the oxygen level in the tissue ({see} Appendix~\ref{sec:oxygenmodel} for details).

\subsection{Potassium buffering by glial cells}
\label{sec:buffer}
Astrocytes, a type of glial cell, play important roles in the instigation and propagation of CSD as well as in neurovascular coupling through neurotransmitter-mediated signaling pathways~\cite{attwell2010glial}.  A principal role of astrocytes is the clearance of local increases {of ECS} potassium~\cite{somjen1975electrophysiology}. This buffering is achieved through a variety of inward rectifying potassium channels in the glial membrane and is bolstered by the extreme polarity of glial cell membranes with membrane potential near the Nernst potential for potassium~\cite{walz2000role}. For this study, we are not interested in the exact mechanisms of glial potassium buffering, but are interested in reproducing accurate potassium dynamics for our model.  Thus, we incorporate astrocyte effects through empirical potassium buffering.
  
 Following Kager et al.~\cite{kager2000simulated}, we modeled the potassium-buffering flux, $v_{\text{buffer}}$, by the following  differential equation, 
 \begin{eqnarray}
 v_{\text{buffer}}(x,t) &=& -\frac{\partial B(x,t)}{\partial t} = \mu_+{[\K]_eB(x,t)}\exp\left(\frac{[\K]_e-5.5}{-1.09} \right) - \mu_- {(B_0 - B(x,t))}, 
 \end{eqnarray}
where $B\ (\mM)$ is the free buffer concentration, the rate constants $\mu_+=\mu_-=8.0\times10^{-6}\textrm{ms}^{-1}$ determine the speed at which potassium is buffered, and $B_0=200 \ \mM$ is the effective total buffer concentration. This equation describes
strong buffering of extracellular potassium for concentrations above $5.5 \ \mM$, but is limited by saturation of the finite buffer. As the amount of buffered potassium increases, re-release of potassium into the extracellular space becomes more favorable. The initial value of the free buffer concentration is set to maintain steady state when the extracellular potassium concentration is at its rest value ($3.5 \ \mM$).
  
\subsection{Neurovascular coupling and oxygen supply}

We now describe how the tissue oxygen level is affected by and influences CSD. First, we assume that there exists an effective blood vessel radius $r$ in the tissue, and that the cerebral blood flow rate (CBF) is given by 
\begin{equation}
\textrm{CBF}=\textrm{CBF}_0\frac{r^4}{r^4_0} \label{effectivebloodradius}
\end{equation}
in which the equilibrium values are again denoted by the subscript 0. This expression is based on the empirical observation that blood flow through the small vessels{,} where nutrient exchange primarily occurs{, can be} modeled as Poiseuille flow where the volume flow rate is proportional to $r^4$~\cite{zamirphysics}. 
      
\begin{figure}[!h]
\begin{center}
\includegraphics[width=19.62pc]{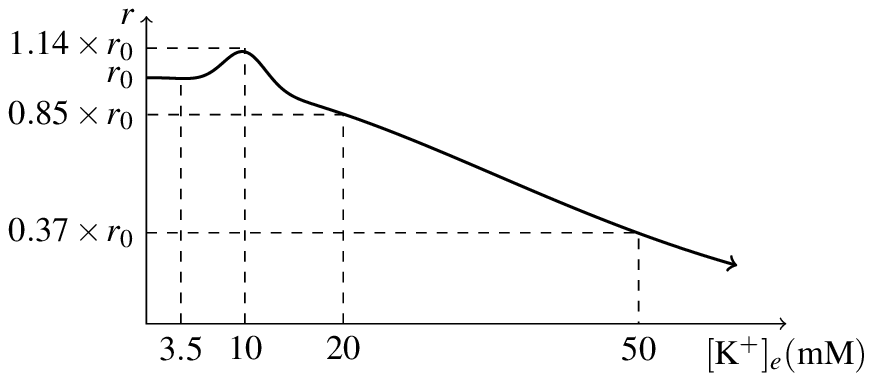}
\end{center}
\caption[{\textbf{Relationship between vascular caliber and $[\K]_e$} Effective vascular radius $r$ as a function of  {ECS} potassium concentration $[\K]_e$.}]{}
\label{fig:radius}
\end{figure}

We use an empirical model for the effective vessel radius $r$, based on replicating the activity observed in many experimental studies on the subject~\cite{knot1996extracellular,mccarron1990potassium}.  Extracellular potassium is known to dilate vessels at lower concentration elevations (less than $\sim17 \ \mM$), and constrict vessels at higher concentration levels.  To mimic this vascular response, Farr and David~\cite{farr2011models} constructed a plot of the radius of cerebral arterioles versus extracellular potassium based upon currents through potassium channels in the membranes of vascular cells.  To reproduce this plot, we  assume that the effective vessel radius is given by
\begin{align}
r([\K]_e)&=r_0{\overbrace{\exp\left\{-\left(\frac{[\K]_e-3.5}{a}\right)^2\right\}}^{\textrm{constriction}}}
\times \frac{\overbrace{\left(1+ be^{-[([\K]_e-10)/c]^2} \right)}^{\textrm{dilation}}}{1+be^{-\left(6.5/c\right)^2}},
\label{eqn:dilation}
\end{align}
which is a product of constricting and dilating terms. Since dendrites are long neuronal processes that occupy a greater volume than cell bodies, we take $[\K]_e$ to be the concentration in the dendritic {ECS}.  The parameter  $a\ (\mM)$ controls the constriction response, $b$ controls the amount of maximal dilation in the vessel, and $c\ (\mM)$ controls the dilation response.  We found values of $a=50 \ \mM$, $b=0.18$, $c=3 \mM$ in (\ref{eqn:dilation}) to mimic the plot of Farr and David given by Fig~\ref{fig:radius}.  We can use this expression in (\ref{effectivebloodradius}) to give us a simple relationship for the CBF in terms of the extracellular potassium concentration.

%

We model the temporal evolution of the tissue oxygen concentration $[\OO]$, using a reaction-diffusion equation
\begin{equation}
\frac{\partial [\OO]}{\partial t}=D_{\OO}\frac{\partial^2 [\OO]}{\partial x^2}+S 
\end{equation}
with nonlinear source term given by
\begin{align} 
\lefteqn{S=\overbrace{\textrm{CBF}\frac{[\OO]_b-[\OO]}{[\OO]_b-[\OO]_0}}^{\textrm{vascular supply}}- \overbrace{\textrm{CBF}_0\times P([\OO])\times (1-\gamma)}^{\textrm{background oxygen consumption}}  } \nonumber \\
&\quad - \underbrace{\textrm{CBF}_0\times P([\OO])\times\gamma \Bigg(\frac{\gamma_{s,\textrm{pump},1}([\K]_{e},[\Na]_{i,s})+\gamma_{d,\textrm{pump},1}([\K]_{e},[\Na]_{i,d})}{\gamma_{\textrm{s,pump},1}([\K]_{e,0},[\Na]_{i,0})+\gamma_{\textrm{d,pump},1}([\K]_{e,0},[\Na]_{i,0}} \Bigg)}_{\textrm{tissue oxygen consumption due to }\Na/\K\textrm{--ATPase}}
\label{eqn:source}
\end{align}
where
\[ P([\OO]) =  \frac{\gamma_{\textrm{pump},2}([\OO])-\gamma_{\textrm{pump},2}(0)}{\gamma_{\textrm{pump},2}([\OO]_0)-\gamma_{\textrm{pump},2}(0)}.
\]
The first term on the right in the source is the amount of oxygen transferred from the blood stream to the tissue and is given by the product of CBF and the normalized concentration difference in oxygen tension between the blood and the tissue. The second term on the right in the source represents the consumption of oxygen by the sodium-potassium exchange pump and other cellular processes that are assumed to remain steady during CSD. The pump consumption is given by the product of the equilibrium CBF and pump rate normalized by the steady-state pump rate. A fraction $0<\gamma<1$ of the total oxygen consumption at steady-state is due to the pump. Experimental estimates of this parameter have ranged from as low as $0.10$~\cite{laughlin1998metabolic} to as high as $0.70$~\cite{aperia2001regulation}. The final term in the source term is the consumption of tissue oxygen beyond their steady state value by $\Na/\K$--ATPase. We include simulations over the full spectrum for thoroughness.  Note that we have $S=0$ at steady-state.

\section{Results\label{sec:simulations}}

We performed one-dimensional simulations by breaking up our model  into a system of ODEs solved using Matlab routine \texttt{ode15s} with reflecting boundary conditions. The computational domain was set to a length of $5.52$ cm and discretized into $46$ grid points. This domain was sufficiently long for the propagating wave to become stable and for boundary effects to remain insignificant.   We used a minimal amount of potassium to induce CSD, finding that for injections of Gaussian boluses with $120$ micron width, a $\K$ concentration of $15 \mM$ was sufficient to induce CSD. This concentration is {the} near previously-reported ceiling for potassium concentration in a non-CSD brain~\cite{heinemann1977ceiling,hansen1988brain} and confirmed to us that our phenomenological potassium buffer was behaving in a physiologically realistic manner.

Throughout these results, we report the duration of the CSD event. We measured the duration of potassium elevation by taking the total amount of time that $[\K]_{e}$ is above $6\ \mM$. The transient sodium channel had negligible effect on the ionic currents during CSD (data not shown), so we chose to omit it from these simulations.

\subsection{Oxygen-clamped simulations}

\begin{figure}[!h]
\begin{center}
\includegraphics[width=19pc]{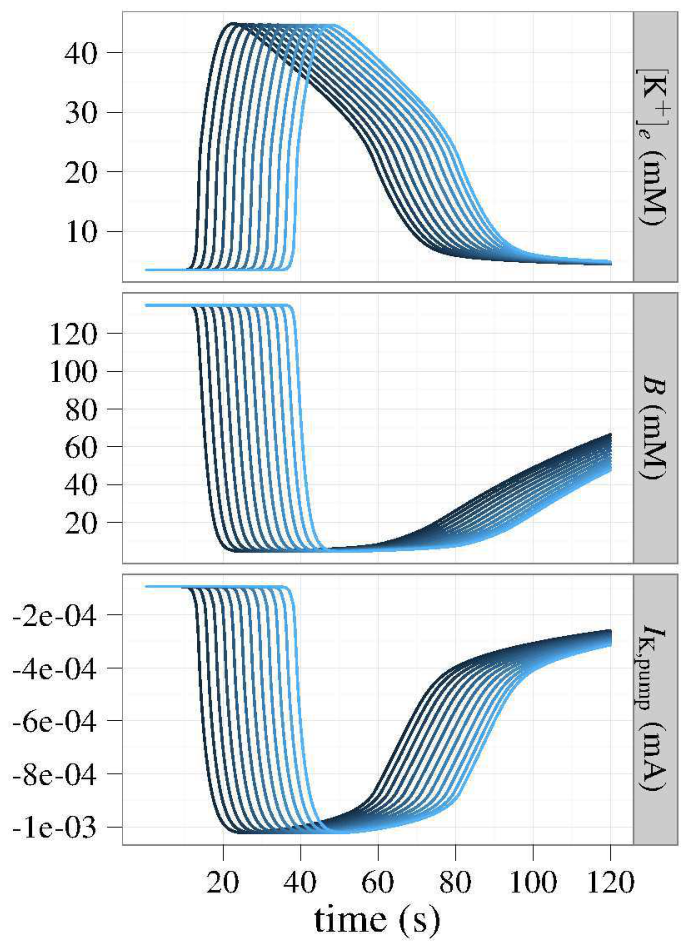}
\end{center}
\caption[{\textbf{CSD in absence of oxygen consumption.} Shown are the time-courses of the propagating waves of  {ECS} potassium concentration, free buffer capacity, membrane potential, and {$\Na/\K$}--ATPase pump rate at positions $120$ microns apart for simulations performed in the absence of oxygen consumption by the $\Na/\K$--ATPase. The $\Na/\K$--ATPase operates at rates determined solely by $[\K]_e$ and $[\Na]_i$.}]{}
\label{fig:uncoupled}
\end{figure}

First, we simulated our model with the oxygen coupling completely turned off, i.e., $\gamma=0$~(see Eq~\ref{eqn:source}).  Thus, the {$\Na/\K$--}ATPase does not consume any oxygen, and we assume that the tissue oxygenation is held at its steady state value. In this situation, we obtain CSD waves with a velocity of $3.2$ millimeters per minute and with {ECS} potassium concentration increasing to a maximum value of $45.7 \ \mM$. This result is similar to {ECS} elevations reported elsewhere in the literature~\cite{hansen2008extracellular}. The total duration of the elevated {ECS} potassium concentration is $65.4$ seconds (Fig~\ref{fig:uncoupled}).

The potassium buffer acts similar in that in Kager et al.~\cite{kager2000simulated}. It saturates rapidly, after which it is responsible for a net re-release of potassium into the extracellular environment. The buffer acts this way in all simulations as it is independent of the oxygen level in our model. For this reason, we omit further mention of the potassium buffer.

\subsection{Blood vessel-clamped simulations}

Next, we simulated our theoretical CSD where the effective blood vessel diameter remained fixed, thereby fixing the maximal oxygen flux rate into the tissue.  We varied the oxygen coupling parameter $\gamma$ between $0$ (no oxygen consumption by pump) and $1$ (maximal oxygen coupling) in increments of $0.025$.

\begin{figure}[!h]
\begin{center}
\includegraphics[width=19.62pc]{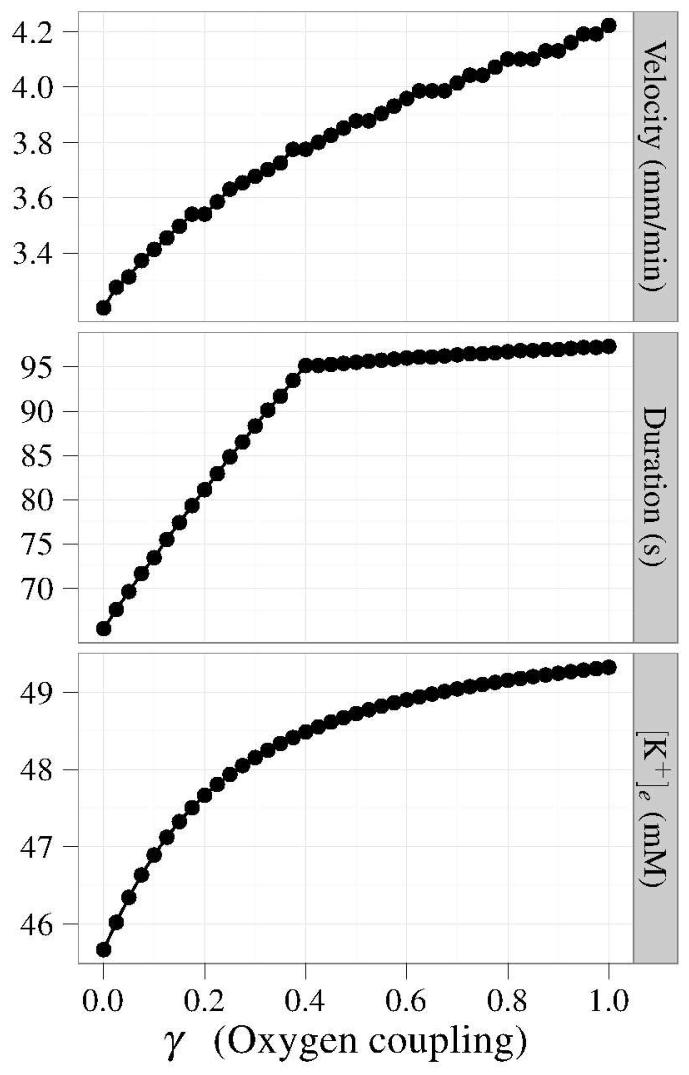}
\end{center}
\caption[{\textbf{Oxygen coupled to $\Na/\K$--ATPase. } Fixing the effective vascular radius while coupling oxygen to the {$\Na/\K$}--ATPase gives a purely consumption-based view of oxygen during CSD. Defining coupling as the fraction of oxygen consumed by the pump at steady-state, these simulations show that all of the CSD characteristics -- speed, duration, maximum $[\K]$ -- increase with an increase in the coupling constant $\gamma$. The duration of the wave increases linearly from $\gamma=0$ to $\gamma\approx0.4$ before increasing linearly at a different rate when $\gamma>0.4$}]{}
\label{fig:fixedvessel}
\end{figure}

\begin{figure}[!h]
\begin{center}
\includegraphics[width=19pc]{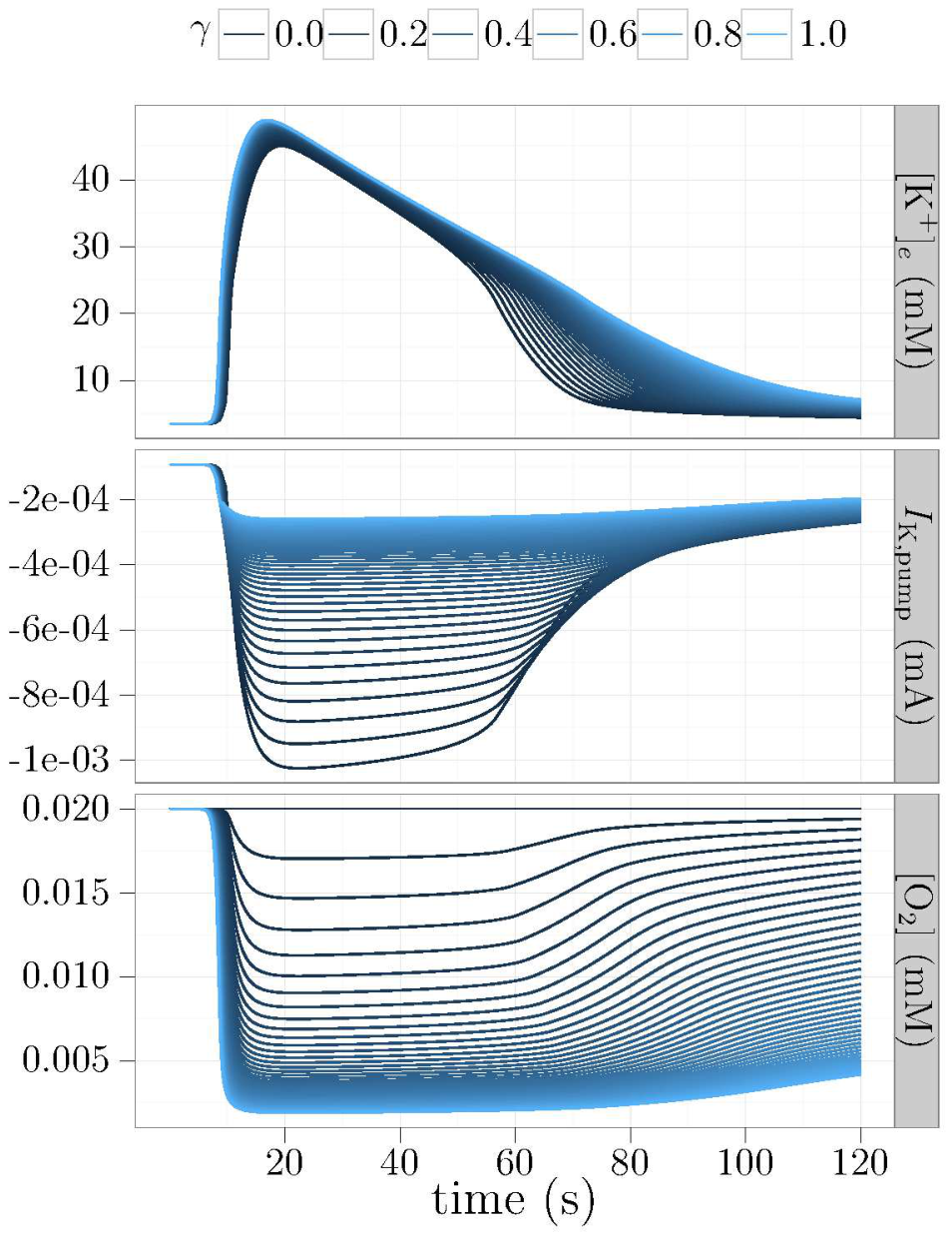}
\end{center}
\caption[{\textbf{Fixed vascular caliber.} Time courses at a fixed position, $780 \mu\textrm{m}$  downstream of the original stimulus. These simulations show the effects of oxygen consumption on CSD for a range of the oxygen coupling parameters, $\gamma$. Increasing $\gamma$ prolongs the duration and magnifies the amplitude of the CSD because it implies that the pump consumes more of the available oxygen, thereby resulting in larger oxygen depletions.}]{}
\label{fig:fixedvessel2}
\end{figure}

The results shown in Fig~\ref{fig:fixedvessel} illustrate the response of CSD to oxygen coupling. By increasing $\gamma$, we increase the duration and the amplitude of the CSD waveform. To get a deeper understanding of these results, Fig~\ref{fig:fixedvessel2} shows the macroscopic observables.  One sees that increases in $\gamma$ lead to greater drops in oxygenation. This deoxygenation results in decreases in the magnitude of the sodium-potassium--ATPase pump current, resulting in longer tails of recovery for both membrane potential and extracellular potassium concentration.

\subsection{Coupled vasculature}

From the previous results, one sees that oxygenation has a definite impact on susceptibility of the tissue to CSD, as defined by its wave propagation speed, and on its recovery.  The next question is whether vasoconstriction has an impact on CSD.
We performed a series of simulations where we activated the vascular coupling that we defined in Eqs.~\ref{eqn:dilation} and \ref{eqn:source}.  We simulated an idealized vascular response, using parameters mimicking the vascular behavior of Farr and David~\cite{farr2011models}. In these simulations, we varied the oxygen coupling parameter, $\gamma$, in order to see the variability in behavior one might expect to see in an idealized CSD experiment. 

\begin{figure}[!h]
\begin{center}
\includegraphics[width=19.62pc]{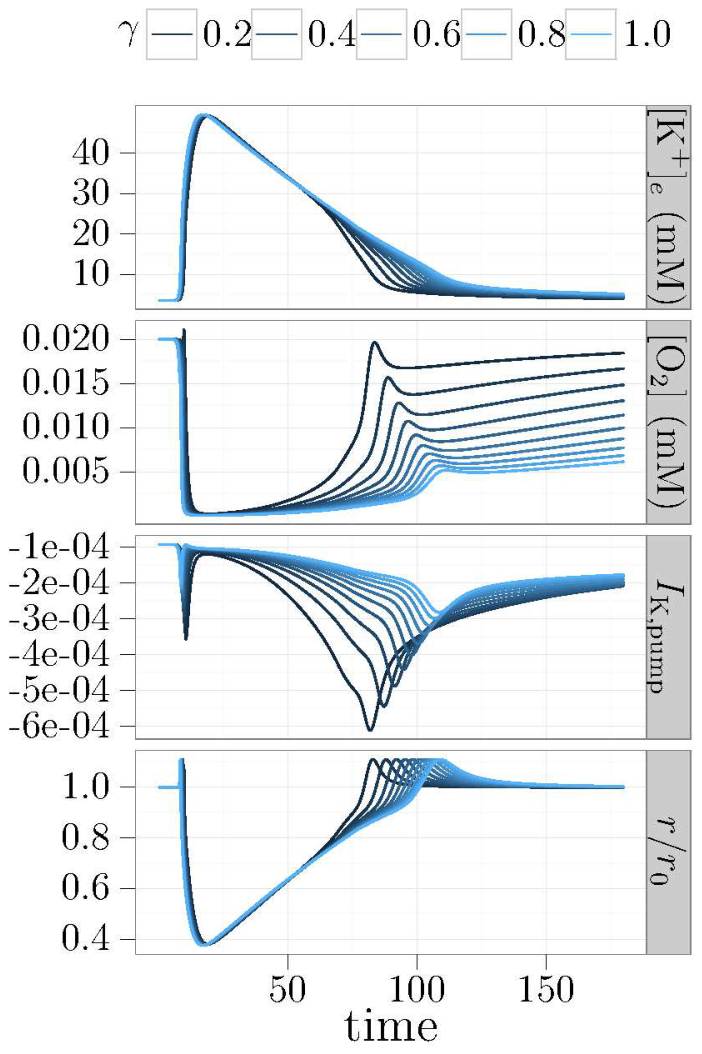}
\end{center}
\caption[{\textbf{``Typical" vascular coupling.}  Time courses at a fixed position, $780 \mu\textrm{m}$ downstream of the original stimulus. Simulations performed using vessel of Farr and David~\cite{farr2011models}. After a short dilation period, the vessels constrict significantly to about 40\% of their rest radii, before recovering.}]{}
\label{fig:farrdavidcsd}
\end{figure}

The results, shown in Fig~\ref{fig:farrdavidcsd}, illustrate the effects of vascular coupling. One can see that the effective blood vessel radius drops rapidly to approximately 40\% of its original value. {This} drop leads to steep sustained reductions in the oxygen concentration during the metabolic challenge {from} the CSD wave. The deoxygenation is reflected in the pump rate, where a large increase in inward potassium current at the beginning of CSD is rapidly diminished as oxygen is depleted. The implications of this chain of events are visible in the {ECS} potassium profile.

\begin{figure*}[!h]
\begin{center}
\includegraphics[width=41pc]{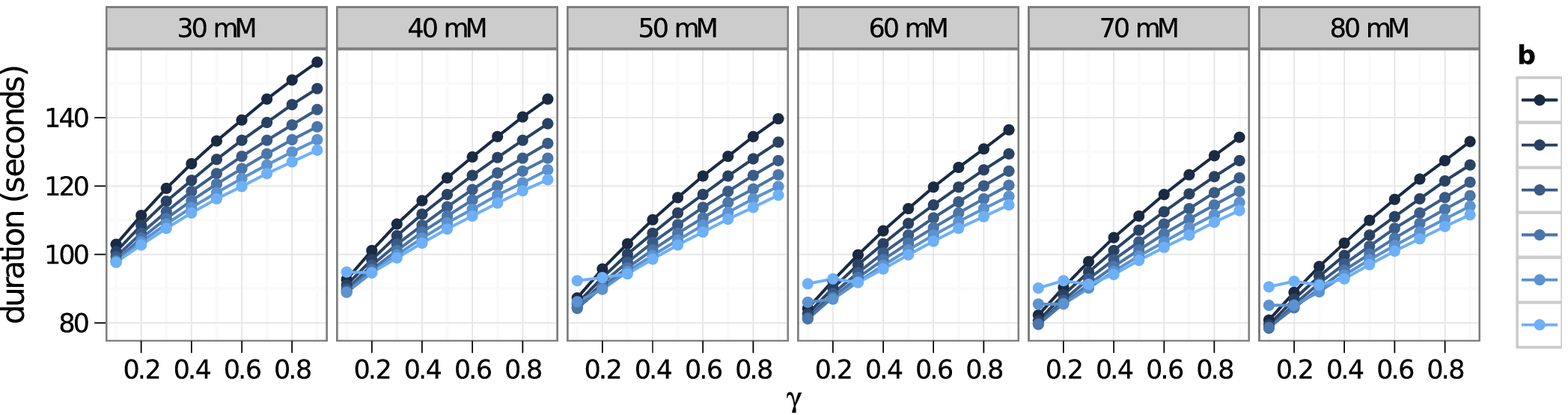}
\end{center}
\caption[{\textbf{Duration of CSD for different vascular responses.} Shown are CSD durations plotted against percent dilation $b$ (the maximum vasodilation), constriction parameter $a$ (this parameter has the units of mM and is the width of a Gaussian curve that controls how fast $r$ drops as $[\K]$ increases~(Eq.~\ref{eqn:dilation})), and oxygen coupling constant $\gamma$. Duration is defined by the length of time that potassium concentration is elevated to a level greater than $6 \ \mM$. Increasing $a$, (shown from $30-80 \ \mM$) decreases the amount of constriction, resulting in quicker recovery from CSD. Likewise, increasing $b$, which increases the maximum dilation of the vessels, also reduces the duration of CSD.}]{}
\label{fig:unfixedduration}
\end{figure*}

\begin{figure*}[!h]
\begin{center}
\includegraphics[width=41pc]{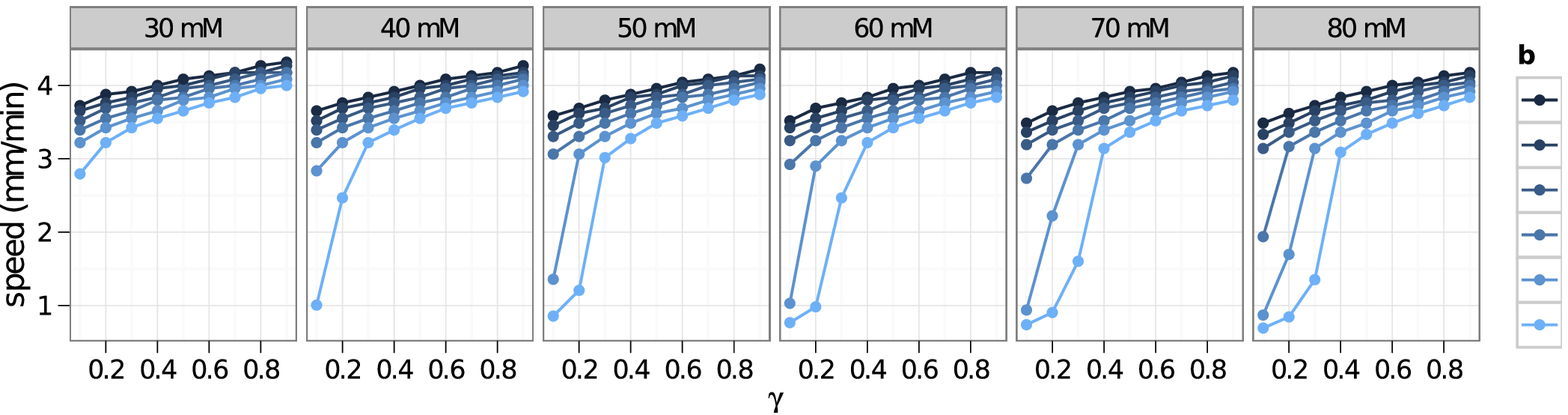}
\end{center}
\caption[{\textbf{Speed of CSD for different vascular responses.} Increasing the constriction parameter $a$ makes the vessel less sensitive to potassium, weakening the resulting constriction. The speed decreases as $a$ increases from $30 \ \mM$ up to $80 \ \mM$, but increasing the amount of dilation by increasing $b$ seems to have a greater effect.}]{}
\label{fig:unfixedspd}
\end{figure*}

As we noted earlier, {the} vascular response is variable {and depends} on factors such {as} species, {individuals}, and metabolic states. In Fig~\ref{fig:unfixedduration}, the effects of this variability are seen in simulations across a grid of values for the parameters $a,b,\gamma$~(Eq.~\ref{eqn:dilation}).  The duration of the CSD event decreases as the parameter $a$ (ECS $\K$ concentration at 63\% constriction) and the parameter $b$ (percent maximal dilation) {are} increased.  In Fig~\ref{fig:unfixedspd}, we have plotted the speed of CSD waves across {a} spectrum of possible vascular responses. These vascular responses are parameterized mainly by {the} two parameters, {$a$ and $b$}.



\section{Discussion\label{sec:discussion}}

 Since its discovery, CSD has been known to involve massive changes in vascular caliber, and hence, perfusion~\cite{le1944spreading}. Up to now, these {vascular} changes, which can have profound effects on cortical function and thus on CSD itself, have not been incorporated into CSD models. Using our model, we were able to examine the effects of vascular activity on CSD. 

\subsection{Relationship between vascular activity and CSD}

 Our numerical experiments confirmed that oxygen delivery plays a significant role in the dynamics of CSD. Even in the absence of constriction, oxygen depletion is seen. Additionally, vasoconstriction and vasodilation were seen to further modify the characteristics of the CSD wave, particularly in the low-$\gamma$ regime. By varying the relationship between vasoconstriction and {ECS} potassium, we achieved a continuum of CSD responses that could explain real physiological variability across species, within members of the same species, and {for}  the same animal. 

\subsubsection{CSD uncoupled from vascular activity}

Our modeling of CSD uncoupled from oxygen concentration resembles current CSD models, which rely on similar assumptions and conductances. Such models are relevant to CSD in {{\it ex vivo}} brain slices, where the perfusing medium has contact with an essentially unlimited source of oxygen. In this case, the pump is always able to perform at a rate determined by $[\K]_e$ and $[\Na]_i$, and the total pump current is constrained only by the total number of available pumps present, provided that there is sufficient nutrient available. Our model predicts that CSD waves {propagate} more slowly and recover more quickly in slices than {{\it in vivo}}.

\subsubsection{CSD coupled with vascular activity}

Coupling an inert vasculature {with} CSD, one obtains a sense of the metabolic demands that CSD places on the cortex. The large displacements of sodium and potassium require the use of energy to {return the cortex to homeostasis.}  {The} physical constraints placed on blood delivery by the physiologically reasonable assumption of finite volume-flow-rate were seen to result in depletion of oxygen and reduction in potassium flux through the pumps. This finding is consistent with {{\it in vivo}} measurements of tissue oxygen and mitochondrial redox state -- even with mild dilation. Several studies have found that both variables moved into a more-reduced oxidative state~\cite{takano2007cortical,sakadvzic2009simultaneous}. In our model, the depletion of oxygen is due to increased metabolic demand driven by the {$\Na/\K$}--ATPase.

\subsubsection{The effects of vasoconstriction/vasodilation}

Furthermore, we show that vasoconstriction can both decrease the tissue's ability to slow down the CSD wave, and impair its ability to recover. These effects are due to the reductions in blood flow causing a significant additional drop in cortical oxygenation levels. In Fig~\ref{fig:farrdavidcsd}, the vessel is seen to constrict to approximately 40\% {of} its original radius. This constriction is within the experimental range reported in Chang et al.~\cite{chang2010biphasic}. Due to the {power law} relationship between blood flow and effective blood vessel radius, even a small reduction in effective radius has large blood flow implications. A 60\% constriction results in blood flow dropping to 2.6\% of its original value. This effect is visible in the oxygenation level, which undergoes further decreases. Our simulations show that both wave speed and recovery time increase when {vascular caliber is reduced}.

Our simulations also show that vasodilation plays a role in CSD. Predictably, vasodilation appears to precondition the tissue so that it is  {better} able to withstand the increase of  {ECS} $\K$ that accompanies the CSD wave. Due to the power law relationship, an increase of 14\% in the vessel radius results in a 70\% increase in blood volume flow rate.  The result shown in Fig~\ref{fig:unfixedspd} is a tissue that is {better} able to withstand potassium elevations, thereby causing slow-downs in the CSD wave. Vasodilation (beyond the effective resting radius) seems to play a significant role in the recovery from CSD, causing decreases in the recovery time (Fig~\ref{fig:unfixedduration}).

\subsection{Clinical and translational implications}

If a major goal of CSD modeling is to understand its role in human disease, incorporation of perfusion and metabolism is an essential step. CSD is a near-complete leveling of ionic gradients which challenges the ability of homeostatic mechanisms to compensate. CSD is also triggered by changes in perfusion {and} metabolism. Finally, during CSD, the relationships between brain activity, {perfusion,} and metabolism - neurovascular coupling - are altered. These complex relationships can be expected to lead to a variety of responses, depending on the state of the brain tissue surrounding the depolarization. Biological data from animals and humans bears out this complexity and variability. Spreading depolarizations can be relatively innocuous  - repetitive CSD in mouse over several days appears to cause no overt injury~\cite{sukhotinsky2011chronic}. However, they can also be quite harmful, enlarging infarct and contusion areas in both animals and humans~\cite{dreier2009cortical,dreier2011role}. These deleterious effects are almost certainly due to alterations in the vascular response to tissue depolarization, and thus cannot be understood from a modeling standpoint without explicit incorporation of perfusion and metabolism. 

Our model more realistically represents conditions observed in the brain during experimental CSD and the spreading depolarizations of migraine and brain injury. Though it simplifies a complex vascular/metabolic response, it has the distinct advantage of making specific, quantifiable predictions which can be used to generate hypotheses for further experimentation. A particular advantage is the ability to explore the whole ``CSD/metabolic parameter space," which is not possible in biological experiments. This modeling could have important implications for study of {the role of} CSD in migraine, as the conditions which could generate such a massive depolarization in an awake behaving person remain obscure.

\subsection{Assumptions, limitations, and future directions}
In this study, in order to make our model as widely applicable as possible, we have not considered the geometry of any particular vascular network. Our lumped model is assumed to be a good approximation of oxygen delivery dynamics in the brain, in general, for gray-matter.  Further insight into the system may be gained by targeting specific tissue types for simulation.

Our simplified model relating arterial effective diameter to $\K$ is likely applicable during CSD but does not mechanistically explain all the subtleties of neurovascular coupling under more normal conditions. Though $\K$ is involved in the coupling of neural and astrocytic activity to blood flow~\cite{filosa2004calcium}, other mediators including arachidonic acid derivatives, purines, nitric oxide, and possibly neurotransmitters are involved as well~\cite{brennan2010update,attwell2010glial,gordon2008brain}. For the purposes of CSD, however,   {the} supra-physiological swings in  {ECS} potassium likely {provide} a good leading-order approximation of vascular behavior. Many of these other mediators also work via their effects on potassium channels~\cite{farr2011models}.

Swelling of individual cells~\cite{mazel2002changes,kager2000simulated,shapiro2001osmotic} and the tissue as a whole~\cite{chang2010biphasic} appears to occur during CSD. Whether cell-swelling plays an important  role in CSD is a controversial topic, as recent computational studies have cast some doubt on its impact~\cite{yao2010continuum,bennett2008quantitative}. For this reason, we have omitted cell swelling from our model.

Finally, this model is not equipped to account for the complexity of the post-CSD state. Chang et al.~\cite{chang2010biphasic} found that extracellular $\K$ levels remained constant during a long--lasting hypoperfusion and depolarization that was found to follow CSD. As this phase may be clinically relevant, future modeling will focus on trying to understand the $\K$--independent mechanisms involved in the etiology of this period. 

\section*{Acknowledgements}
The authors would like to thank Tom Chou (UCLA Biomathematics and Mathematics)  and Tim David (University of Canterbury) for their insightful comments. 

The work presented here is based on work and discussions held by the authors before, during, and after the American Institute of Mathematics (AIM) SQuaRE workshops on Modeling Cortical Spreading Depression held in Palo Alto, California in May 2010, August 2011, and August 2012.   We thank AIM for their financial support of the SQuaRE {workshops} and special thanks go to Estelle Basor for her warm hospitality and ensuring a convenient and stimulating work environment.  

We also thank the Banff International Research Station for Mathematical Innovation and Discovery (BIRS) who supported a Focussed Research Group on Cerebral Blood Flow, Neurovascular Coupling, and Cortical Spreading Depression in June 2013.

The authors thank  grant number T32GM008185 from
the National Institute of General Medical Sciences and grant  number 58386MA from the Army Research Office (JCC); the National Institutes of
Health (NINDS NS059072 and NS070084) and the Department of Defense (CDMRP PR100060)  (KCB); the Natural Sciences and Engineering Research Council of Canada and the Mathematics in Technology and Complex Systems (HH), the National Science
Foundation through grant DMS-1022848 (RMM, HH, JJW), and agreement No. 0635561 (JCC), and the Research Grants Council of the Hong Kong Special Administrative Region, China, CityU 104211 (JJW).

\onecolumn
\appendix

%
%
%
%
%
%
%

\section{Cross-Membrane Currents and Parameter Values}\label{appendix} 
We assume that the membrane currents in the soma used by Kager et al.~\cite{kager2000simulated} apply here.  The total cross-membrane current in the soma is given by the sum of the active and passive sodium, potassium, chloride, and nonspecific ionic currents. The sodium current is
$I_{s,\textrm{Na,tot}}=I_{s,\textrm{Na,P}}+I_{s,\textrm{Na,Leak}}+I_{s,\textrm{Na,Pump}}$ where ${I_{s,\textrm{Na,P}}}$ is the persistent sodium current, $I_{s,\textrm{Na,leak}}$ is the sodium leak current, and $I_{s,\textrm{Na,pump}}$ is the {sodium} current through the pump. {Note that we have removed $I_{s,\textrm{Na,T}}$, the fast transient sodium current, since this current was shown in~\cite{yao2010continuum} to not make any fundamental difference in the way CSD propagates.} The active and passive potassium currents are $I_{s,\textrm{K,tot}}=I_{s,\textrm{K,DR}}+I_{s,\textrm{K,A}}+I_{s,\textrm{K,leak}}+I_{s,\textrm{K,pump}}$ where $I_{s,\textrm{K,DR}}$ is the potassium delayed rectifier current, $I_{s,\textrm{K,A}}$ is the transient potassium current,  {$I_{s,\textrm{K,leak}}$ is the potassium leak current, and} $I_{s,\textrm{K,pump}}$ is the potassium current through the pump. 
The passive chloride leak current is $I_{s,\textrm{leak}}$.  In the dendritic compartment, the NMDA currents, $I_{d,\textrm{Na,NMDA}}$ and $I_{d,\textrm{K,NMDA}}$, are added to the total current. 
The mathematical expressions for these channels do not differ across the dendritic and somatic compartments so we omit the compartment prefix ($s,d$) from this point forward.


The cross-membrane currents are modeled using the Goldman-Hodgkin-Katz (GHK) formulas for the active membrane currents given by
\begin{align}
{I_{\textrm{ion},\textrm{GHK}}}=
m^ph^q\frac{g_{\textrm{ion,GHK}}F E_m\left[[\textrm{ion}]_i-\exp\left(-\frac{E_m}{\phi}\right)[\textrm{ion}]_e\right]}{\phi\left[1-\exp\left(-\frac{E_m}{\phi}\right)\right]},
\end{align} 
where the permeability is absorbed into the parameter $g_{\textrm{ion,GHK}}$. The factors in the parameter $\phi=RT/F$ are $R{=8.31}$ {mV coulomb/mmol K}, the universal gas constant, $T=310$ K, the absolute temperature, and $F{=96.485}$ {coulomb/mmol}, the Faraday constant. 
The GHK equation is suitable when there is a large difference in concentrations between the ICS and ECS compartments, as argued by Koch and Segev~\cite{keener2009mathematical}.
For the passive leak currents we used the Hodgkin-Huxley (HH) model  given by
\begin{eqnarray} 
I_{ion,HH}&=&g_{ion,HH}\left(E_m-E_{\textrm{ion}}\right).
\end{eqnarray}
In these general expressions for the GHK and HH types of currents, $g_{\textrm{ion,GHK}}$ and $g_{\textrm{ion,HH}}$ are the conductances associated with the channels for $\textrm{ion}=\Na$ and $\K$, and $m$ and $h$ are the activation and inactivation gating variables, respectively, for the different GHK-modeled channels that are ion-specific.  Note that the $g_{\textrm{ion,HH}}$ conductances for the sodium and potassium leak currents and the $g_{\textrm{leak}}$ conductance for the chloride leak current are assumed to be constant.  The variables, $E_{\textrm{ion}}$, are the Nernst potentials for $\text{ion} = \Na\ \text{and } \K$ given by
\begin{equation}
E_{\textrm{ion}}=\phi\log\frac{[\textrm{ion}]_e}{[\textrm{ion}]_i}.
\end{equation}
For the chloride leak current, the equivalent Nernst potential is $-70 \mV$.  

The gating variables, $m$ and $h$, satisfy the following relaxation equations
\begin{eqnarray}
\frac{dm}{dt}&=&\alpha_m(1-m)-\beta_m m,\\
\frac{dh}{dt}&=&\alpha_h(1-h)-\beta_h h,
\end{eqnarray}
and the values of $\alpha$ and $\beta$ are given in Table~\ref{parameter}, along with the exponents $p$ and $q$~\cite{kager2002conditions}.  The extracellular volume is assumed to be $15\%$ of the intracellular volume, i.e., $V_e=0.15V_i$.

We use the following procedure to choose the initial (equilibrium) values for the gating variables and ion concentrations. We first set the membrane potential at $E_m=-70$ mV and the sodium and potassium concentrations (listed in Table~\ref{InitialH}), from which we compute the parameters $\alpha$ and $\beta$ (using the formulas in Table~\ref{parameter}). We then compute the equilibrium values of $m$ and $h$ as
\[m=\frac{\alpha_m}{\alpha_m+\beta_m},\quad h=\frac{\alpha_h}{\alpha_h+\beta_h}.\]
Next we choose the leak conductances, $g_{\textrm{Na,leak}}$ and $g_{\textrm{K,leak}}$, by setting 
\[I_{\textrm{Na,tot}}=I_{\textrm{K,tot}}=0.\]
Finally, we determine the chloride leak conductance in $I_{\textrm{leak}}$ by assuming that $g_{\textrm{leak}}=10g_{\textrm{Na,leak}}$. 
 
 {The initial resting ion concentrations consistent with those in~\cite{kager2000simulated,kager2002conditions} are obtained by modifying the  {$\Na/\K$} exchange pump function and running the model equations until a steady state is reached. These values are given in Table~\ref{InitialH} together with other parameter values.}

\section{Derivation of oxygen-dependent model for the  {$\Na/\K$--}ATPase}\label{sec:oxygenmodel}

The {$\Na/\K$}--ATPase is a transmembrane protein with two extracellular binding sites for potassium, three internal binding sites for sodium, and a single intracellular binding site for ATP. Its activity is {well described} by a sixth-order Hill-like equation of the form
\begin{equation}
I_{\text{pump}} = I_{\text{max}}\left(1+\frac{[\K]_{e,0}}{[\K]_e}\right)^{-2}\left(1+\frac{[\Na]_{i,0}}{[\Na]_i}\right)^{-3}\left(1+\frac{[\ATP]_{i,0}}{[\ATP]}\right)^{-1}.
\end{equation}
To use this formulation for the pump activity, one needs to understand ATP dynamics within the cell. Fortunately,
the biochemical pathways through which ATP is generated are well known~\cite{cloutier2009integrative,heinrich1996regulation}. Unfortunately, these pathways are
complicated, involving a panoply of  intermediate products and enzyme-substrate interactions. The full details of
the biochemical pathway, however, are not relevant towards achieving an understanding of ATP availability
during CSD. The authors in~\cite{cloutier2009integrative,heinrich1996regulation} provide an ATP mass-balance equation that is a sum of aerobic pathways,
anaerobic pathways, and consumption by pumps. It takes the form
\begin{equation}
\frac{\partial[\ATP]}{\partial t} = \left(\underbrace{v_{\text{mito}}}_\textrm{aerobic}+\underbrace{v_{\text{CK}}+v_{\text{PGK}} -v_{\text{HK}}-v_{\text{PK}}}_{\textrm{anaerobic processes}}-v_{\text{pump}} \right)\left(1 - \frac{d\AMP}{d\ATP}\right)^{-1},
\end{equation}
where $v_\textrm{mito}$ is oxygen dependent and takes the form
\[
v_\textrm{mito}\propto\left({1+\frac{[\OO]_0}{[\OO]}}\right)^{-1}.
\]

These studies   {focus on} $\ATP$:$\AMP$ ratios in the regime where the reaction catalyzed {by} the adenylate kinase enzyme is fast relative to other reactions. In neurons, we do not expect the adenylate kinase reactions to dominate due to the presence of aerobic processes.

Instead, we looked at the system of equations presented in these studies to find the behavior when the system is dominated by aerobic production of {ATP} and consumption of ATP. In this regime, the inter-conversion between ADP and ATP dominates over the conversion of AMP to ADP, so that $d\textrm{AMP}/d\textrm{ATP} \ll1$. We also note that the oxygen-dependent ATP-production by the mitochondria ($v_{\text{mito}}$) is {a} fast reaction, thereby yielding the  quasi-steady-state relationship between ATP and $\OO$, $\ATP=\textrm{const}_1+\textrm{const}_2\left(\frac{\OO}{\textrm{const}_3+\OO}\right)$, where $\textrm{const}_1$ refers to anaerobic ATP production.

 \section{Parameter Tables}\label{sec:params}
\begin{Table}[h]
\begin{center}
\caption{Parameter values for active membrane ionic currents, from~\cite{kager2000simulated,kager2002conditions}. Units are given in Table 2.}
 \label{parameter}
\vspace{12pt}
\begin{tabular}{|c|c|c|l|}   \hline
Currents & $g_{ion,GHK}$ & Gates & Voltage-Dependent Rate Constants \\
mA/cm$^2$ & mA cm & $m^ph^q$ & \\ \hline
$I_{\textrm{Na,P}}$ & 2 $\times 10^{-6}$ & $m^2h$ & \large $\alpha_m =
\frac{1}{6\left(1+\exp[-(0.143E_m+5.67)]\right)}$ \\[.1in]
& & & \large $\beta_m =
\frac{\exp[-(0.143E_m+5.67)]}{6\left(1+\exp[-(0.143E_m+5.67)]\right)}$ \\
[.1in] & & & \large $\alpha_h =5.12 \times 10^{-8}{\exp[-(0.056E_m+
2.94)]}$ \\ [.1in] & & & \large $\beta_h =  \frac{1.6
\times10^{-6}}{1+\exp[-(0.2E_m+8]}$ \\ [.1in] \hline 
$I_{\textrm{K,DR}}$ & 10 $\times 10^{-5}$ & $m^2$ & \large $\alpha_m = 0.016
\frac{E_m+34.9}{1-\exp[-(0.2E_m+6.98)]}$ \\ [.1in] & & & \large
$\beta_m = \small 0.25 \exp[-(0.25E_m+1.25)]$ \\ [.1in] \hline
$I_{\textrm{K,A}}$ & 1 $\times 10^{-5}$ & $m^2h$ & \large $\alpha_m = 0.02
\frac{E_m\mV^{-1}+56.9}{1-\exp[-(0.1E_m+5.69)]}$ \\ [.1in] & & & \large
$\beta_m = 0.0175\frac{E_m+29.9}{\exp(0.1E_m+2.99)-1}$ \\ [.1in] & &
& \large $\alpha_h = 0.016 \exp[-(0.056E_m+4.61)]$ \\ [.1in] & & &
\large $\beta_h = \frac{0.5}{1+\exp[-(0.2E_m+11.98)]}$ \\ [.1in]
\hline
 $I_{\textrm{NMDA}}$ & 1 $\times 10^{-5}$ & $mh$ & \large $\alpha_m =
\frac{0.5}{1+\exp\left(\frac{13.5-[\K]_e}{1.42}\right)}$ \\ [.1in]&
& & \large $\beta_m =0.5-\alpha_m$ \\
[.1in] & & & \large $\alpha_h = \frac{1}{2000 \left(1+
\exp\left[\frac{[\K]_e-6.75}{0.71}\right]\right)}$ \\ [.1in]& & &
\large $\beta_h =5\times10^{-5}-\alpha_h$ \\
[.1in] \hline
\end{tabular}
\end{center}
\end{Table}


\begin{Table}[h]
\begin{center}
\caption{{Initial resting values and other relevant parameter values for the computations, following Kager et al.~\cite{kager2000simulated,kager2002conditions}.}}
\label{InitialH}
\vspace{12pt} 
\begin{tabular}{|c|l|l|}   \hline Parameter & Value & Unit \\
\hline $R_a$ (input resistance of dendritic tree) &  $1.83\times10^5 $ &ohms\\
\hline $d_s$ (diameter of soma) & $5.45\times 10^{-4}$ & cm \\
\hline $d_d$ (diameter of dendrite) & $9.39\times10^{-5}$ & cm\\
\hline $\delta_d$ (half-length of dendrite) & $4.5\times10^{-2}$ & cm\\
\hline $A_s$ (surface area of soma)& $1.586\times10^{-5}$ & cm$^2$\\
\hline $A_d$ (surface area of dendrite) &  $2.6732\times10^{-4}$ & cm$^2$\\
\hline $V_s$ (volume of soma) & $ 2.160\times10^{-9}$ & cm$^3$ \\
\hline $V_d$ (volume of dendrite) & $5.614\times10^{-9}$ & cm$^3$\\
\hline $C_m$(membrane capacitance) &$7.5\times 10^{-5}$ & s /$\Omega$
cm$^2$\\
\hline  $I_{max}$ ($\Na/\K$--ATPase rate)&$1.48\times10^{-3}$ & mA / cm$^{2}$\\
\hline $E_m$ &-70 & mV\\
\hline $[\K]_e$  &{ 3.5} & mM \\
\hline $[\K]_i$ &{  133.5}& mM \\
\hline $[\Na]_e$ &{ 140}& mM \\
\hline $[\Na]_i$&{10}& mM\\
\hline $[\OO]_0$ & $2\times10^{-2}$ & mM\\
\hline $\text{CBF}_0$ & $2.5\times10^{-2}$ & $\mM/s$ \\
\hline $D_{\OO}$ & $5\times10^{-4}$ & cm$^2$/s\\
\hline $[\OO]_b$ & $4\times10^{-2}$ & mM\\
\hline
\end{tabular}
\end{center}
\end{Table}
\newpage
\clearpage
 \bibliography{references}

\begin{thebibliography}{10}
\providecommand{\url}[1]{\texttt{#1}}
\providecommand{\urlprefix}{URL }
\expandafter\ifx\csname urlstyle\endcsname\relax
  \providecommand{\doi}[1]{doi:\discretionary{}{}{}#1}\else
  \providecommand{\doi}{doi:\discretionary{}{}{}\begingroup
  \urlstyle{rm}\Url}\fi
\providecommand{\bibAnnoteFile}[1]{%
  \IfFileExists{#1}{\begin{quotation}\noindent\textsc{Key:} #1\\
  \textsc{Annotation:}\ \input{#1}\end{quotation}}{}}
\providecommand{\bibAnnote}[2]{%
  \begin{quotation}\noindent\textsc{Key:} #1\\
  \textsc{Annotation:}\ #2\end{quotation}}
\providecommand{\eprint}[2][]{\url{#2}}

\bibitem{bures1974mechanism}
Bures J, Buresova O, Krivanek J (1974) The mechanism and applications of Leao's
  spreading depression of EEG activity.
\newblock Academic.
\bibAnnoteFile{bures1974mechanism}

\bibitem{kraig2002spreading}
Kraig R, Kunkler P (2002) Spreading depression: a teleological means for
  self-protection from brain ischemia.
\newblock In: Cerebrovascular Disease (22nd Princeton Research Conference). pp.
  142--157.
\bibAnnoteFile{kraig2002spreading}

\bibitem{tfelt-hansen2010migraine-csd}
Tfelt-Hansen P (2010) History of migraine with aura and cortical spreading
  depression from 1941 and onwards.
\newblock Cephalalgia 30: 780--792.
\bibAnnoteFile{tfelt-hansen2010migraine-csd}

\bibitem{hadjikhani2001mechanisms}
Hadjikhani N, Sanchez~del Rio M, Wu O, Schwartz D, Bakker D, et~al. (2001)
  Mechanisms of migraine aura revealed by functional {MRI} in human visual
  cortex.
\newblock Proceedings of the National Academy of Sciences 98: 4687-4692.
\bibAnnoteFile{hadjikhani2001mechanisms}

\bibitem{lashley1941patterns}
Lashley K (1941) Patterns of cerebral integration indicated by the scotomas of
  migraine.
\newblock Archives of Neurology and Psychiatry 46: 331-339.
\bibAnnoteFile{lashley1941patterns}

\bibitem{le1944spreading}
Le{\~a}o A (1944) Spreading depression of activity in the cerebral cortex.
\newblock Journal of Neurophysiology 7: 359--390.
\bibAnnoteFile{le1944spreading}

\bibitem{brennan2010update}
Brennan K, Charles A (2010) An update on the blood vessel in migraine.
\newblock Current Opinion in Neurology 23: 266-274.
\bibAnnoteFile{brennan2010update}

\bibitem{charles2009cortical}
Charles A, Brennan K (2010) Cortical spreading depression - new insights and
  persistent questions.
\newblock Cephalalgia 29: 1115--1124.
\bibAnnoteFile{charles2009cortical}

\bibitem{busija2008mechanisms}
Busija D, Bari F, Domoki F, Horiguchi T, Shimizu K (2008) Mechanisms involved
  in the cerebrovascular dilator effects of cortical spreading depression.
\newblock Progress in Neurobiology 86: 417--433.
\bibAnnoteFile{busija2008mechanisms}

\bibitem{chang2010biphasic}
Chang J, Shook L, Biag J, Nguyen E, Toga A, et~al. (2010) Biphasic direct
  current shift, haemoglobin desaturation and neurovascular uncoupling in
  cortical spreading depression.
\newblock Brain 133: 996-1012.
\bibAnnoteFile{chang2010biphasic}

\bibitem{piilgaard2009persistent}
Piilgaard H, Lauritzen M (2009) Persistent increase in oxygen consumption and
  impaired neurovascular coupling after spreading depression in rat neocortex.
\newblock Journal of Cerebral Blood Flow \& Metabolism 29: 1517--1527.
\bibAnnoteFile{piilgaard2009persistent}

\bibitem{somjen2001mechanisms}
Somjen G (2001) Mechanisms of spreading depression and hypoxic spreading
  depression-like depolarization.
\newblock Physiological Reviews 81: 1065--1096.
\bibAnnoteFile{somjen2001mechanisms}

\bibitem{lauritzen2010clinical}
Lauritzen M, Dreier J, Fabricius M, Hartings J, Graf R, et~al. (2010) Clinical
  relevance of cortical spreading depression in neurological disorders:
  migraine, malignant stroke, subarachnoid and intracranial hemorrhage, and
  traumatic brain injury.
\newblock Journal of Cerebral Blood Flow \& Metabolism 31: 17--35.
\bibAnnoteFile{lauritzen2010clinical}

\bibitem{dreier2011role}
Dreier J (2011) The role of spreading depression, spreading depolarization and
  spreading ischemia in neurological disease.
\newblock Nature Medicine 17: 439--447.
\bibAnnoteFile{dreier2011role}

\bibitem{sukhotinsky2008hypoxia}
Sukhotinsky I, Dilekoz E, Moskowitz M, Ayata C (2008) Hypoxia and hypotension
  transform the blood flow response to cortical spreading depression from
  hyperemia into hypoperfusion in the rat.
\newblock J Cerebral Blood Flow \& Metabolism 28: 1369--1376.
\bibAnnoteFile{sukhotinsky2008hypoxia}

\bibitem{tuckwell1978mathematical}
Tuckwell H, Miura R (1978) A mathematical model for spreading cortical
  depression.
\newblock Biophysical Journal 23: 257--276.
\bibAnnoteFile{tuckwell1978mathematical}

\bibitem{kager2000simulated}
Kager H, Wadman W, Somjen G (2000) Simulated seizures and spreading depression
  in a neuron model incorporating interstitial space and ion concentrations.
\newblock Journal of Neurophysiology 84: 495-512.
\bibAnnoteFile{kager2000simulated}

\bibitem{shapiro2001osmotic}
Shapiro B (2001) Osmotic forces and gap junctions in spreading depression: a
  computational model.
\newblock Journal of Computational Neuroscience 10: 99--120.
\bibAnnoteFile{shapiro2001osmotic}

\bibitem{yao2010continuum}
Yao W, Huang H, Miura R (2010) A continuum neuronal model for the instigation
  and propagation of cortical spreading depression.
\newblock Bulletin of Mathematical Biology 73: 2773-2790.
\bibAnnoteFile{yao2010continuum}

\bibitem{somjen2008computer}
Somjen G, Kager H, Wadman W (2008) Computer simulations of neuron-glia
  interactions mediated by ion flux.
\newblock Journal of Computational Neuroscience 25: 349--365.
\bibAnnoteFile{somjen2008computer}

\bibitem{kager2002conditions}
Kager H, Wadman W, Somjen G (2002) Conditions for the triggering of spreading
  depression studied with computer simulations.
\newblock Journal of Neurophysiology 88: 2700-2712.
\bibAnnoteFile{kager2002conditions}

\bibitem{takano2007cortical}
Takano T, Tian G, Peng W, Lou N, Lovatt D, et~al. (2007) Cortical spreading
  depression causes and coincides with tissue hypoxia.
\newblock Nature Neuroscience 10: 754--762.
\bibAnnoteFile{takano2007cortical}

\bibitem{andrew2007physiological}
Andrew RD, Labron MW, Boehnke SE, Carnduff L, Kirov SA (2007) Physiological
  evidence that pyramidal neurons lack functional water channels.
\newblock Cerebral Cortex 17: 787--802.
\bibAnnoteFile{andrew2007physiological}

\bibitem{nicholson1981ion}
Nicholson C, Phillips J (1981) Ion diffusion modified by tortuosity and volume
  fraction in the extracellular microenvironment of the rat cerebellum.
\newblock The Journal of Physiology 321: 225--257.
\bibAnnoteFile{nicholson1981ion}

\bibitem{nicholson2001diffusion}
Nicholson C (2001) Diffusion and related transport mechanisms in brain tissue.
\newblock Reports on Progress in Physics 64: 815-884.
\bibAnnoteFile{nicholson2001diffusion}

\bibitem{keener2009mathematical}
Keener J, Sneyd J (2009) Mathematical Physiology: I: Cellular Physiology.
\newblock Springer Verlag.
\bibAnnoteFile{keener2009mathematical}

\bibitem{klein2003principles}
Klein J, Ergin M (2003) Principles of Cerebral Protection during Operations on
  the Thoracic Aorta.
\newblock BC Decker, second edition, 291--303 pp.
\bibAnnoteFile{klein2003principles}

\bibitem{chih2003energy}
Chih C, Roberts E (2003) Energy substrates for neurons during neural activity:
  A critical review of the astrocyte-neuron lactate shuttle hypothesis.
\newblock Journal of Cerebral Blood Flow \& Metabolism 23: 1263--1281.
\bibAnnoteFile{chih2003energy}

\bibitem{attwell2010glial}
Attwell D, Buchan A, Charpak S, Lauritzen M, MacVicar B, et~al. (2010) Glial
  and neuronal control of brain blood flow.
\newblock Nature 468: 232--243.
\bibAnnoteFile{attwell2010glial}

\bibitem{somjen1975electrophysiology}
Somjen G (1975) Electrophysiology of neuroglia.
\newblock Annual Review of Physiology 37: 163--190.
\bibAnnoteFile{somjen1975electrophysiology}

\bibitem{walz2000role}
Walz W (2000) Role of astrocytes in the clearance of excess extracellular
  potassium.
\newblock Neurochemistry International 36: 291--300.
\bibAnnoteFile{walz2000role}

\bibitem{zamirphysics}
Zamir M (2000) The Physics of Pulsatile Flow.
\newblock Springer-Verlag, Berlin and New York.
\bibAnnoteFile{zamirphysics}

\bibitem{knot1996extracellular}
Knot H, Zimmermann P, Nelson M (1996) Extracellular {K}$^+$-induced
  hyperpolarizations and dilatations of rat coronary and cerebral arteries
  involve inward rectifier {K}$^+$ channels.
\newblock The Journal of Physiology 492: 419-430.
\bibAnnoteFile{knot1996extracellular}

\bibitem{mccarron1990potassium}
McCarron J, Halpern W (1990) Potassium dilates rat cerebral arteries by two
  independent mechanisms.
\newblock American Journal of Physiology (Heart and Circulatory Physiology 28)
  259: H902--H908.
\bibAnnoteFile{mccarron1990potassium}

\bibitem{farr2011models}
Farr H, David T (2011) Models of neurovascular coupling via potassium and {EET}
  signalling.
\newblock Journal of Theoretical Biology 286: 13-23.
\bibAnnoteFile{farr2011models}

\bibitem{laughlin1998metabolic}
Laughlin S, van Steveninck R, Anderson J (1998) The metabolic cost of neural
  information.
\newblock Nature Neuroscience 1: 36--41.
\bibAnnoteFile{laughlin1998metabolic}

\bibitem{aperia2001regulation}
Aperia A (2001) Regulation of sodium/potassium {ATP}ase activity: impact on
  salt balance and vascular contractility.
\newblock Current Hypertension Reports 3: 165--171.
\bibAnnoteFile{aperia2001regulation}

\bibitem{heinemann1977ceiling}
Heinemann U, Dieter~Lux H (1977) Ceiling of stimulus induced rises in
  extracellular potassium concentration in the cerebral cortex of cat.
\newblock Brain Research 120: 231--249.
\bibAnnoteFile{heinemann1977ceiling}

\bibitem{hansen1988brain}
Hansen AJ, Nedergaard M (1988) Brain ion homeostasis in cerebral ischemia.
\newblock Neurochemical Pathology 9: 195--209.
\bibAnnoteFile{hansen1988brain}

\bibitem{hansen2008extracellular}
Hansen A, Zeuthen T (2008) Extracellular ion concentrations during spreading
  depression and ischemia in the rat brain cortex.
\newblock Acta Physiologica Scandinavica 113: 437--445.
\bibAnnoteFile{hansen2008extracellular}

\bibitem{sakadvzic2009simultaneous}
Sakad{\v{z}}i{\'c} S, Yuan S, Dilekoz E, Ruvinskaya S, Vinogradov S, et~al.
  (2009) Simultaneous imaging of cerebral partial pressure of oxygen and blood
  flow during functional activation and cortical spreading depression.
\newblock Applied Optics 48: D169--D177.
\bibAnnoteFile{sakadvzic2009simultaneous}

\bibitem{sukhotinsky2011chronic}
Sukhotinsky I, Dilekoz E, Wang Y, Qin T, Eikermann-Haerter K, et~al. (2011)
  Chronic daily cortical spreading depressions suppress spreading depression
  susceptibility.
\newblock Cephalalgia 31: 1601--1608.
\bibAnnoteFile{sukhotinsky2011chronic}

\bibitem{dreier2009cortical}
Dreier J, Major S, Manning A, Woitzik J, Drenckhahn C, et~al. (2009) Cortical
  spreading ischaemia is a novel process involved in ischaemic damage in
  patients with aneurysmal subarachnoid haemorrhage.
\newblock Brain 132: 1866--1881.
\bibAnnoteFile{dreier2009cortical}

\bibitem{filosa2004calcium}
Filosa J, Bonev A, Nelson M (2004) Calcium dynamics in cortical astrocytes and
  arterioles during neurovascular coupling.
\newblock Circulation Research 95: e73--e81.
\bibAnnoteFile{filosa2004calcium}

\bibitem{gordon2008brain}
Gordon G, Choi H, Rungta R, Ellis-Davies G, MacVicar B (2008) Brain metabolism
  dictates the polarity of astrocyte control over arterioles.
\newblock Nature 456: 745--749.
\bibAnnoteFile{gordon2008brain}

\bibitem{mazel2002changes}
Mazel T, Richter F, Vargov{\'a} L, Sykov{\'a} E (2002) Changes in extracellular
  space volume and geometry induced by cortical spreading depression in
  immature and adult rats.
\newblock Physiol Res 51: S85--S93.
\bibAnnoteFile{mazel2002changes}

\bibitem{bennett2008quantitative}
Bennett M, Farnell L, Gibson W (2008) A quantitative model of cortical
  spreading depression due to purinergic and gap-junction transmission in
  astrocyte networks.
\newblock Biophysical Journal 95: 5648--5660.
\bibAnnoteFile{bennett2008quantitative}

\bibitem{cloutier2009integrative}
Cloutier M, Bolger F, Lowry J, Wellstead P (2009) An integrative dynamic model
  of brain energy metabolism using in vivo neurochemical measurements.
\newblock Journal of Computational Neuroscience 27: 391--414.
\bibAnnoteFile{cloutier2009integrative}

\bibitem{heinrich1996regulation}
Heinrich R, Schuster S (1996) The Regulation of Cellular Systems.
\newblock Springer.
\bibAnnoteFile{heinrich1996regulation}

\end{thebibliography}
\newpage
\clearpage
\renewcommand\listfigurename{Figure Legends}
\listoffigures

\end{document}